\documentclass[aps,pra,showpacs,twocolumn]{revtex4}

\usepackage{graphicx}
\makeatletter
\g@addto@macro{\UrlBreaks}{\UrlOrds}
\makeatother

\begin{document}
\preprint{APS/123-QED}

\title{Optimising the signal-to-noise ratio in measurement of photon pairs with detector arrays.}

\author{Eric Lantz}
\author{Paul-Antoine Moreau}
\author{Fabrice Devaux}

\affiliation{Institut FEMTO-ST  D\'epartement d'Optique PM Duffieux
\\UMR CNRS - Universit\'e de Franche-Comt\'e $n\,^{\circ}\mathrm{ 6174}$,
Route de Gray  25030 Besan\c con Cedex, FRANCE}

\date{\today}

\begin{abstract}
To evidence multimode spatial entanglement of spontaneous down-conversion, detector arrays allow a full field measurement, without any a priori selection of the paired photons. We show by comparing results of the recent literature that electron-multiplying CCD (EMCCD) cameras allow, in the present state of technology, the detection of quantum correlations with the best signal-to-noise ratio (SNR), while intensified CCD (ICCD) cameras allow at best to identify pairs. The SNR appears to be proportional to the square root of the number of coherence cells in each image, or Schmidt number. Then, corrected estimates are derived for extended coherence cells and not very low and not space-stationary photon fluxes. Finally, experimental measurements of the SNR confirm our model.
\end{abstract}


\maketitle

\section{Introduction}

 Entanglement in continuous spatial variables has attracted great attention during the recent years because of its potentially very high dimensionality. Indeed, for spontaneous parametric down-conversion (SPDC) the dimensionality, or Schmidt number ~\cite{Law2004,Exter2006,Devaux2012}, corresponds to the number of resolution cells, or transverse spatial modes, where the size of a resolution cell scales as the inverse of the phase matching angular range ~\cite {Devaux95}. This number can exceed several thousands in the transverse plane. A full characterization of the quantum properties of photons in this huge space would require a precise determination of arrival times for each resolution cell, which is clearly beyond the present technology capacities. One has therefore to choose between two less ambitious objectives:

 - first, few photons are picked at two localized places (pixels) that correspond to photon pairs with a reasonably high probability, by using a temporal coincidence circuitry \cite{Howell2004, Sullivan2005, Dixon2012}. Characterization of spatial entanglement is made by measuring an excess of coincidences for certain couples of pixels. Though useful to characterize spatial entanglement in view of its use in quantum information protocols, where we need to know that two detected photons do form a pair, these methods are questionable for demonstrating basic properties like Einstein-Podolsky-Rosen (EPR) paradox \cite{Einstein1935}, because of the detection of a very few part of the incident photons with an a priori selection criterion. For example, it can be demonstrated that a classical single beam can show an apparent violation of the Heisenberg uncertainty principle if measurements are made in one dimension: see Appendix. Moreover, measurements must be repeated for each couple of pixels, which is extremely time-consuming. Attempts to reduce this time have been recently performed by using compressive-sensing \cite{Howland2013}, however at the expense of the precise spatial location of the pairs, just as in the second type of methods described in the following.

 -Second, large bi-dimensional detector arrays allow a full-field detection of all arriving photons with an equal chance: some photons are not detected because of limited quantum efficiency or false thresholding and some detection events do not correspond to actual photons because of noise, but these imperfections do not lie on a priori assumptions about the nature of the light, except its weakness allowing a single-photon sensitivity. Intensified CCD (ICCD) cameras \cite{Jost1998, Haderka2005, Oemrawsingh2002} or more recently electron-multiplying CCD (EMCCD) cameras \cite{Blanchet2008,Blanchet2010,Moreau2012,Edgar2013} allow this ultimate sensitivity. The advantages and drawbacks of these two types of cameras can be roughly described as follows. ICCD have a poor quantum efficiency (50$\%$ at most) and a quite extended spatial impulse response but may have a very low noise level for a short exposure time. Conversely,  EMCCDs may have a very high quantum efficiency and an impulse response almost limited to a single pixel, but are noisy. The basic reason of this difference is that the noisy transfer characterizing all CCDs occurs before amplification in EMCCDs, resulting in amplification of the clock-induced charge (CIC) \cite{Lantz2008}, while CIC is completely negligible as compared to the intensified signal in an ICCD.

 The goal of this paper is to discuss how to optimize the use of detector arrays for measuring spatial quantum correlations. Tasca et al. have already proposed in a recent paper \cite{Tasca2013} an optimization criterion lying on the maximization of the visibility of these correlations. While this criterion can be pertinent if looking for detection of actual pairs, we show that evidencing quantum correlations on a whole image is at best performed by maximizing the signal-to-noise ratio (SNR), where signal refers here to the correlation peak due to twin photons and noise refers to accidental coincidences as well as the camera noise.

 The paper is organized as follows. In section 2, we develop a simple model of imaging that allows us to discuss the pertinence of both criteria and their correspondence with an optimized mean photon flux level. Section 3 is devoted to estimation problems in a realistic situation. In section 4, we show and compare with theory an experimental curve of SNR versus the intensity level. Then we conclude.

\section{Idealized model}

 Let us consider an idealized type II SPDC experiment at very low gain, where the signal and idler beams are detected on two different cameras, or equivalently on two non overlapping zones of a single camera. In the spirit of this oversimplified model, we suppose that diffraction is negligible: for each photon that is impinging on the pixel $S_{i}$ on the signal detector, a photon is impinging on the "twin" pixel $I_{i}$ on the idler side, where $i$ gives the position of the pixel in the bi-dimensional transverse plane with a system of coordinates ensuring the same $i$ for twin photons in the near-field as well in the far-field. A straightforward derivation from basic principles gives \cite{Brambilla2001} the probability $P_{ijmn}$ of detecting two photons signal and idler on the pixels $i$ and $j$ and in the temporal modes $m$ and $n$:
 
 \begin{equation}\label{Pijmn}
 P_{ijmn}=\eta^{2}(v^{4}+\delta_{ij}\delta_{mn}u^{2}v^{2})
 \end{equation}
where the amplitude gains $u$ and $v$ obey the unitarity condition $u^{2}-v^{2}=1$, $\delta$ is the Kroenecker symbol and $\eta$ is the quantum efficiency, that takes into account the probability of no detection of the photon as well as the probability of a peak generated by a photoelectron of height below the threshold level. Let $\mathcal{M}$ be the number of temporal modes where SPDC occurs. $\mathcal{M}$ can be estimated as the ratio between the bandwidth of the SPDC and the spectral width of the pump beam, typically $10^{3}$ in our experiments, for a Fourier-tranform pump with a pulse duration of several hundredths of picoseconds. By summing the elementary probabilities given by Eq.\ref{Pijmn} on the $\mathcal{M}^{2}$ couples of temporal modes, we obtain the mean number of coïncidences $P_{ij}$ between the the pixels $i$ and $j$:
\begin{eqnarray}\label{Pij}
P_{ij}&=&\eta^{2}(\mathcal{M}^{2}v^{4}+\delta_{ij}\mathcal{M}u^{2}v^{2})\nonumber\\
&=& m^{2}+\delta_{ij}(\frac{m^{2}}{\mathcal{M}}+\eta m)
\end{eqnarray}	
	
Where $m=\eta \mathcal{M} v^{2}$ is the mean number of photoelectrons recorded on one pixel. By subtracting the $m^{2}$ term, Eq.\ref{Pij} is in agreement with the covariances given in \cite{Brida2011}. If m is sufficiently small to neglect the weak probability of multiple detections on a single pixel, $P_{ij}$ can be considered as the probability of a coincident detection of two signal and idler photons. The influence of multiple detections will be considered in subsection \ref{satur}, by using a relation valid for on-off detectors \cite{Blanchet2008}. Because of the great number of temporal modes and of the low value of the fluence $m$, the term $\frac{m^{2}}{\mathcal{M}}$ is extremely weak and can be neglected. See also some comments about this term at the end of this section. The first term on the right hand side of Eq.\ref{Pij} appears as a probability of accidental coincidences, i.e. the probability of coincidences if the signal and the idler beam would be independent. Note however that this term exists also, with the same value, for "twin" modes, where all photons arrive by pairs. For mathematical convenience, we will call this term "accidental coincidences" even in this latter case.  We suppose now that a spurious detection due to the detector noise occurs with a low probability $p_{n}$. The probability of detection of a "false pair" that is due either to accidental coincidences between photons or between  photon and noise or between noise and noise is given  $Pacc=(m+p_{n})^{2}$.  In a practical experiment, it means that the result $G_{ij}$ of a numerical product, for each couple of pixels, of the signal and the idler values evidences for $i=j$ a correlation peak of height proportional to $Pacc+\eta m$, surrounded by a continuous background of mean value proportional to $Pacc$. In analogy with optical images, the visibility $V$ of the correlation peak can be computed as:

 \begin{eqnarray}\label{V}
V&=&\frac{G_{max}-G_{min}}{G_{max}+G_{min}}=\frac{\eta m}{\eta m+2(m+p_{n})^{2}}
\end{eqnarray}

Maximizing the visibility clearly implies working with $m$ as low as possible inasmuch as the rare incident photons overcome the detector noise \cite{Tasca2013}. A straightforward derivation leads to a maximization of $V$ for $m=p_{n}$.

  Note that with such a criterion, ICCD cameras may exhibit superiority over EMCCDs: the maximum visibility becomes equal to

\begin{equation}\label{Vmax}
V_{max}=\frac{\frac{\eta}{p_n}}{8+\frac{\eta}{p_n}}
\end{equation}

which is an increasing function of $\frac{\eta}{p_{n}}$. Maximizing $\frac{\eta}{p_{n}}$ has a direct physical interpretation. Indeed, for $m\simeq p_{n}$, the ratio of true pairs detection over the false ones becomes proportional to $\frac{\eta}{p_{n}}$. Hence, ICCDs appear as working at best to identify true pairs. Despite their lower quantum efficiency, low exposure times can reduce their dark noise to a negligible level, inducing a very low detector noise because of the absence of CIC and a threshold level well above the readout noise. In ref \cite{Oemrawsingh2002}, noise is primary due to stray light. They obtained 230 true pairs of photons for 1839 coincidences recorded on 436000 images. True pairs of photons can be separated from the false ones because they form a correlation peak as expected, though not on a single pixel like in our idealized model.

 On the other hand, stating that two given recorded  photons form a true pair with high probability seems difficult with an EMCCD, even a posteriori like in the preceding example, because of CIC.  Nevertheless, we argue in the following  that EMCCDs efficiently allow evidencing quantum correlations like a degree of EPR paradox. Efficiently means here with averages on a minimum of independent pixels and images. We will see that using a single couple of images (one couple for position and one for momentum for EPR demonstration) does not seem out of reach. A correlation peak is evidenced if it cannot be confounded, with high probability,  with random fluctuations of the background noise. Hence, we have to prove that the true correlation peak is much higher than the highest fluctuation of the correlation background. The magnitude of this highest fluctuation is proportional to the standard deviation of the estimates of the correlation background. We give in the following the simplest computation of this standard deviation, in the frame of our idealized model. After the recording of K images, $G_{ij}$ is estimated for each couple of pixels as:

 \begin{equation}\label{Gij}
\widehat{G_{ij}}=\frac{1}{K}\sum_{k=1}^{K}N_{i,k}N_{j,k}
\end{equation}

where $N_{i,k}$ ($N_{j,k}$) $=0$ or $1$ are the binary thresholded intensities on the image k and the hat stands, here and in all the following, as "estimator of". For pixels that do not correspond to true pairs  ($i\neq j$), the variance $\sigma^{2}_{G}$ of $\widehat{G_{ij}}$ is given by :

\begin{eqnarray}\label{varacc}
\sigma^{2}_{G}&=&\frac{1}{K} \left(Pacc(1-Pacc)^{2}+(1-Pacc)(0-Pacc)^{2}\right)\nonumber\\
&=&\frac{1}{K}Pacc(1-Pacc)
\end{eqnarray}

The signal-to-noise ratio ($SNR$) is therefore given by:
\begin{equation}\label{SN}
SNR=\frac{P_{ii}-Pacc}{\sigma_{G}}\simeq \sqrt{K}\frac{\eta\  m}{m+p_{n}}
\end{equation}

where we have assumed $Pacc\ll 1$. In this expression, the signal is defined as the height of the "twin" peak above the noise floor of mean $Pacc=(m+p_{n})^{2}$.  Because of the absence of squaring in the denominator, the SNR increases with $m$, unlike the visibility, as long as $m$ is not much greater than $p_{n}$. For $m\gg p_{n}$, the SNR becomes proportional to the quantum efficiency. Hence EMMCD's exhibit a clear superiority versus ICCD's when optimizing the SNR is the relevant criterion. However, a too large $m$ would result in a too large probability of  recording two photoelectrons on the same pixel. Though results are corrected in order to take into account this probability (see subsection \ref {satur}), it seems safe to work at a level minimizing the total probability of all false detections, either of no photoelectron where there is one, or of one photoelectron when there is zero or more than one. We  have shown \cite {Lantz2008} that this level corresponds to $m=0.15$ photoelectrons/pixel, giving a value of $\frac{\eta\  m}{m+p_{n}} \approx0.85$, for $p_{n}\approx10^{-2}$ and $\eta>0.9$. This should be compared for an ICCD to $\frac{\eta\  m}{m+p_{n}} \approx\eta\approx0.5$ for an ICCD of third generation~\cite{Andor}. Only for very low fluxes, $m<10^{-2}$, the negligible value of $p_{n}$ for an ICCD will lead to a better SNR than for an EMCCD.\\
The second message of Eq.(\ref{SN}) is the increasing of the $SNR$ with the number of acquired images. Moreover the space-stationarity assumption allows a further improvement: all couples of pixels in an image that have the same difference of spatial positions possess the same statistical properties. We follow now closely our previous work ~\cite{Devaux2012} to show that the normalized intercorrelation function between both images gives directly the degree of their quantum correlation. For a detection of a photon S at $\overrightarrow{r_{S}}$, the
probability density of detection of a photon I at
$\overrightarrow{r_{I}}$ can be written as:
\begin{equation}\label{p1}
p(\overrightarrow{r_{I}}|\overrightarrow{r_{S}})=p(\overrightarrow{r_{I}})+f(\Delta
r)
\end{equation}
where p$(\overrightarrow{r_{I}})=m+p_{n}$ is the probability density of
detection of a photelectron issued from another pair or from noise (accidental coincidences)
and $f(\Delta r)$ is the probability density of detection of the
twin photon, with $\Delta r
=\|\overrightarrow{r_{2}}\pm\overrightarrow{r_{1}}\|$, $+$ holding
for the far-field (correlation of momenta on opposite modes) and
$-$ for the near-field. It is assumed translational invariance, circular symmetry and independence of the pairs (pure
SPDC without further amplification). Hence, if N$_S$ is the number
of photons S detected on a surface S$_S$ and N$_I$ the
corresponding quantity for photons I on a surface S$_I$, we have:
\begin{eqnarray}\label{n1n2}
<N_S N_I>&=&\int_{S_S}dr_{S}^{2}\int_{S_I}dr_{I}^{2}\
p(\overrightarrow{r_{S}}\, and \,\overrightarrow{r_{I}})\\
&=&\int_{S_S}dr_{S}^{2}\int_{S_I}dr_{I}^{2}\left\{
p(\overrightarrow{r_{S}}) p(\overrightarrow{r_{I}})+
p(\overrightarrow{r_{S}})f(\Delta r)\right\}\nonumber
\end{eqnarray}
Therefore, the probability of detection in S$_I$ of the twin
photon I of the photon S detected on S$_S$ is simply given by:
\begin{equation}\label{fdeltar}
F(S_I)=\int_{S_I}dr_{I}^{2} \ f(\Delta r)=\frac{<N_S N_I>-<N_S><N_I>}{<N_S>}
\end{equation}
If S$_S$ and S$_I$ have the same size, this expression can be
symmetrized and becomes the normalized intercorrelation function:
\begin{equation}\label{fs2}
F(S_I)=F(S_S)=\frac{<N_S N_I>-<N_S><N_I>}{(<N_S>+<N_I>)/2}
\end{equation}
For independent pairs, i.e. a pure Poisson statistics for each beam, this quantity can also be expressed as a
function of the variance of the difference between N$_S$ and
N$_I$:
\begin{eqnarray}\label{fdeltarvmoins}
<N_S>=<N_I>=<(N_S)^2>-<N_S>^2 \Rightarrow\nonumber\\
F(S_I)=1-\frac{<(N_S- N_I)^2>}{<(N_S+N_I)>}
\end{eqnarray}

Hence, if we assume a pure Poisson statistics of the pairs, a positive value of the correlation peak corresponds to a sub-shot-noise level of the variance of the difference, ensuring the demonstration of quantum correlations. The validity of this Poisson assumption is a subtle matter, that is discussed in the next lines. On one hand, even if the statistics for each temporal mode is thermal, the number of photons in one temporal mode $v^{2}$ is much lower than one, ensuring that the variance $u^{2}v^{2} \approx v^{2}$, i.e. the same equality as for a Poisson beam. On the other hand, it has been extensively shown (see for example \cite {Basano2005}) that classical thermal correlations can be evidenced even in this situation. This is the Handbury-Twiss-Brown effect \cite{HBT1953}: if a classical thermal beam is divided in two, the rate of coincidences in a coherence cell, though low, is twice the accidental rate outside this coherence cell. However, this effect is negligible in our situation: $\mathcal{M}\approx 10^{3}$ independent temporal modes add on each pixel and it is very unlikely that two pairs come from the same temporal mode. Quantitatively, the rate of accidental coincidences, $m^{2}$ in Eq.\ref{Pij} is much greater than the classical thermal term $\frac{m^{2}}{\mathcal{M}}$. We have compared in the Fig.3 of \cite{Devaux2012} the experimental estimation of the degree of correlation between images issued either from Eq.\ref{fs2} or \ref{fdeltarvmoins}, with a good agreement. Anyway, because of the possibility of classical correlations for thermal beams that obey a statistics exhibiting some Poisson characters, the only safe criterion ensuring the quantum character of a correlation peak is the further demonstration of the sub-shot-noise level of the variance of the difference:

\begin{equation}\label{vardiff} 
	\frac{<(N_S- N_I)^2>}{<(N_S+N_I)>}<1 
\end{equation}
	
Indeed, for a symmetrical process like SPDC, this sub-shot-noise level is equivalent to the violation of a Cauchy-Schwarz inequality \cite{Brambilla2001,Kheruntsyan2012}. However, we will see in the next section some possible artifacts in the estimation of both the correlation peak and the variance of the difference, that, if not corrected, could make appear as quantum classical or not corelated beams. \\
Because of the translational invariance, the means in eq. \ref
{fs2} and  \ref {fdeltarvmoins} can be estimated by spatial
averages on the different pixels in one image. Hence, we can estimate $P(\mathbf{r_{S}}\ and\ \mathbf{r_{I}})$ by summing the products of the fluences for pixels having the same difference of spatial positions:

\begin{equation}\label{Gdeltar}
\widehat{G(\mathbf{\Delta r})}=\frac{1}{D\ K}\sum_{d=1}^{D}\sum_{k=1}^{K}N_{S,d,k}N_{I,d',k}
\end{equation}

 where D is the total number of pixels in each area and d' is the coordinate in the idler image corresponding to a shift of $\mathbf{\Delta r}$ with respect to the coordinate d in the signal image. Edge effects, i.e. $d'>D$ can be easily treated: we are actually interested by the small values of $\mathbf{\Delta r}$. It is possible to define, for the range of these small values, a region of interest of D pixels on the signal image sufficiently small to be sure that the idler value $N_{I,d',k}$ has been acquired.\\
 The covariance $\widehat{H(\mathbf{\Delta r})}$ between pixels can be computed by subtracting the estimated means:

 \begin{equation}\label{Fdeltar}
\widehat{H(\mathbf{\Delta r})}=\frac{\sum_{k=1}^{K}\sum_{d=1}^{D}\left(N_{S,d,k}-\overline{N_{S,d}}\right)\left(N_{I,d',k}- \overline{N_{I,d'}}\right)}{D\ (K-1)}
\end{equation}

 Where the horizontal bar means an average over the images for a given pixel: $\overline{N_{S,d}} =\frac{1}{K}\sum_{k=1}^{K}N_{S,d,k}$. Unlike G, H vanishes for $f(\mathbf{\Delta r\neq 0})$ and can be easily computed by using a standard correlation routine.  \\
The interest of this sum over different couples of pixels in an image is of course to considerably increase the SNR. Eq. (\ref{SN}) becomes, by still assuming $m+p_{n}\ll 1$:

 \begin{equation}\label{SNR}
SNR\simeq \sqrt{D\ K}\frac{\eta\  m}{m+p_{n}}
\end{equation}

\section{Realistic conditions}

We now consider the influence of more realistic conditions on the SNR. Specifically, the list of treated items includes:\\

- diffraction effects, leading to a probability of incidence of the twin photon on several pixels forming a coherence area\\
- a non negligible probability of incidence of two photons or more on one pixel\\
- a non space-stationary beam, because of its gaussian shape.

The first effect evidently determines the width of the quantum peak but we show below that it is also strongly connected to the effective SNR. The two other effects can lead to classical beams appearing as quantum. Only after their correction, the criterion of a subpoissonian variance becomes effective to characterize the quantum regime.

\subsection{Diffraction and coherence cells}

It can be demonstrated ~\cite{Saleh2000, Devaux2012} that the conditional probability density in the far-field is proportional to the pump amplitude in this plane, and therefore has the same width inasmuch as this width is much smaller than the width of the phase matching function. A dual relation exists in the near-field: the width of the conditional probability function is given in this plane by the inverse of the phase-matching range. Hence in both planes, the number of transverse modes, or resolution cells in Ref. ~\cite {Devaux95}, has been recognized
~\cite{Law2004,Exter2006} as corresponding to the two-photon Schmidt number.  The total conditional probability is given by the integral of the conditional probability density over the coherence cell. Experimentally, a binning (grouping) of the pixel in either the images or the intercorrelation bi-dimensional function allows the retrieval of a peak of maximal height for the quantum correlation function. To roughly quantify the influence of this binning on the SNR, let us suppose that the conditional probability density is constant over a coherence cell of C pixels. The conditional probability $P_{ii}=\eta\  m$ of our idealized model is now shared between these C pixels, with a corresponding division of the signal to noise ratio if no binning (nothing has changed regarding $P_{acc}$):

\begin{equation}\label{SNRnobin}
SNR_{nobin}\simeq \frac{\sqrt{D\ K}}{C}\frac{\eta\  m}{m+p_{n}}
\end{equation}

By binning these C pixels, we recover a twin signal equal to $\eta\  m$, while the variance of the noise is multiplied by C, leading to a SNR:

\begin{equation}\label{SNRbin}
SNR_{bin}\simeq \sqrt{\frac{D}{C}\ K}\frac{\eta\  m}{m+p_{n}}.
\end{equation}

Hence, with an adequate binning, the SNR appears to be proportional to the square root of the number $N_{C}=\frac{D}{C}$ of coherence cells in the image. The same conclusion holds if the experiment has been designed to ensure a size of a physical pixel corresponding to a coherence cell. It corresponds simply to the case $C=1$ in the preceding equations.\\
It is interesting to give an order of magnitude of the minimum total number of coherence cells ensuring a determination without ambiguities of the quantum correlation peak. If we assume that the fluctuations of the accidental coincidences are Gaussian, they never exceed five standard deviations. For $m\gg p_{n}$, it means that no ambiguity is possible if $SNR>5$, or $N_{C}\ K>\left(\frac{5}{\eta}\right)^2$. For a single couple of images, $K=1$, and an overall quantum efficiency of 0.1, it corresponds to 50 coherence cells in each transverse direction. This overall quantum efficiency corresponds in practice to the integral of the normalized correlation signal, as experimentally determined in the far-field ~\cite{Devaux2012} (see section \ref{experiment}). 50 coherence cells correspond to a degree of EPR paradox of 2500 in each direction, while our best recent results are 600 in a direction and 30 in the orthogonal one \cite{Moreau2014b}. Nevertheless, the objective does not seem out of reach, since the overall quantum efficiency could be probably improved.

\subsection{Several photons incident on one pixel}\label{satur}

Experimentally, the use of an on-off detector like an EMCCD in the thresholding regime leads, for a coherent beam, to a measured variance smaller than the measured mean. This phenomenon can be easily explained by taking into account the cases where two photoelectrons or more are
accumulated in the same pixel. If $\mu$ is the true mean number of
photoelectrons accumulated in one pixel, a thresholding procedure
would give, in the absence of false detections, a measured mean m
given by
\begin{equation}
m=1-p(0)=1-exp(-\mu)
\end{equation} where p(0) is the probability of detecting no photoelectron.
The first equality expresses the fact that the thresholding
procedure is unable to distinguish between one and more
photoelectrons on one pixel, while the second equality reflects
the Poisson distribution of photoelectrons. With the same
hypotheses, the measured variance $\sigma_{m}^{2}$ is given by
\begin{equation}\label{correctv}
\sigma_{m}^{2}=m^2 p(0)+(1-m)^2(1-p(0))=m(1-m) \end{equation} Hence, the
measured variance is smaller than the measured mean, because of
the binary detection. Moreover, the variance of the
difference is affected in the same way as the variance by this
effect and the ratio between these variances gives an estimation
of quantum correlations for independent pairs, with a standard quantum limit equal to 2.
However, the more general criterion separating the classical and the quantum world involves the means, like in Eq. \ref{vardiff}. To employ such a criterion, the mean $m$ must be replaced by the corrected mean $m(1-m)$.\\
This correction was first proposed by our group in Ref. \cite{Blanchet2008}. It was rediscovered, seemingly independently, four years later in a somewhat more general context \cite{Sperling2012}.\\
When the mean is estimated on a small number of samples, the estimator $\widehat{m}$ of $m$ fluctuates and the non-biased estimator of the variance for K samples is given by:

\begin{equation}\label{varK}
\widehat{\sigma_{m}^{2}}=\frac{K}{K-1}\widehat{m}(1-\widehat{m}),\ with\ \widehat{m}=\frac{1}{K}\sum_{k=1}^{K}N_{k} \end{equation}

with an indeterminate result for K=1 ($\widehat{m}$=0 or 1).\\

In the presence of detector noise, all the above reasoning remains valid if we replace the true number of photoelectrons $\mu$ by the number of electrons that is read $\mu+p_{n}$. The measured mean m is now given by $m=1-p(0)=1-exp(-(\mu+p_{n}))$. With this new definition of the measured mean, Eq.(\ref{correctv}) remains valid.

\subsection{Non space-stationary beam}\label{nostat}

We have to perform statistics on the whole gaussian beams to take into account the most part of the photon flux in order to make full field measurements. Therefore, the hypothesis of stationary statistics of our idealized model is completely ruled out. At first sight, it does not seem to have important consequences: (co)variances between independent coherence cells add, like means, and summations allow the test of the sub-shot-noise character of the variance of the difference, by simply assuming a constant fluence on the different pixels of each coherence cell. By also assuming that the mean level of SPDC does not fluctuate from an image to another, we can indeed calculate the  variance of the difference and the corrected mean for each pixel on the K images, and then perform spatial averages. The test of the quantum regime is obtained as:

\begin{equation}\label{vmoinspix}
\frac{(K-1)\sum_{d=1}^{D}\sum_{k=1}^{K}\left(N_{S,d,k}-N_{I,d',k}\right)^2}
{K^{2}\sum_{d=1}^{D}\left(\overline{N_{S,d}}(1-\overline{N_{S,d}})+\overline{N_{I,d'}}(1-\overline{N_{I,d'}})\right)}<1
\end{equation}

On the other hand, a difficulty appears if the corrected mean is calculated image per image, in order to test the quantum regime for each individual image. In this spirit, we define for the image k a coefficient $r_k$  \cite{Blanchet2010}:

\begin{equation}\label{r}
r_k=\frac{\frac{1}{D}\sum_{d=1}^{D}\left(N_{S,d,k}-N_{I,d',k}\right)^2}
{\widehat{N_{S,k}} (1-\widehat{N_{S,k}})+\widehat{N_{I,k}} (1-\widehat{N_{I,k}})}
\end{equation}

where the estimation is performed by an average on the whole image: $\widehat{N_{S,k}}=\frac{1}{D}\sum_{d=1}^{D}N_{S,d,k}$. However, $r_k<1$ for an image, or $\overline{r}<1$ for the average on K images, are not correct criteria of the quantum regime for a SPDC beam with gaussian shape. Indeed, the variance estimate $\widehat{N_{S,k}} (1-\widehat{N_{S,k}})$ includes a deterministic term due to the variation in space of the mean SPDC intensity. More precisely, for a pixel $d$ localized at a given point in the gaussian beam, the measured variance averaged on a great number of realizations with the same experimental conditions as in the image k can be written as:
\begin{equation}\label{truevariance}
\sigma_{m}^{2}=<N_{S,d,k}(1-N_{S,d,k})>
\end{equation}
 In practice, $\sigma_{m}^{2}$ can be estimated by averaging over the pixels of an image as:
 \begin{eqnarray}\label{truemeanvariance}
\widehat{\sigma_{m}^{2}}&=&\frac{1}{D}\sum_{d=1}^{D}<N_{S,d,k}(1-N_{S,d,k})>\nonumber\\
&=& \widehat{N_{S,k}}-\frac{1}{D}\sum_{d=1}^{D} <N_{S,d,k}^{2}>\nonumber\\
&=&<\widehat{N_{S,k}} (1-\widehat{N_{S,k}})>-\frac{1}{D}\sum_{d=1}^{D}<\Delta N_d^2>
\end{eqnarray}
where $\Delta N_d=<N_{S,d,k}-\widehat{N_{S,k}}>$ is the deterministic deviation from the mean of the measured SPDC at the pixel d.\\
 To conclude, the corrected mean overestimates the variance of the random fluctuations because this corrected mean includes also a term of variance due to the deterministic variations between pixels. It can be shown in the same way that the usual estimator of the variance overestimates the random part of the variance of the same quantity for the same reasons. On the other hand, there is no deterministic part in the variance of the difference inasmuch as the deterministic profiles of the signal and the idler are identical, because the mean of this difference is zero whatever the pixels. As a consequence, even decorrelated beams can exhibit an apparent subpoissonian behavior. We have indeed verified on  non correlated experimental images that the variance of the difference appears smaller than the sum of the variances as well than the sum of the corrected means. After subtraction of the deterministic term, the expected equality is restored. This analysis has similarities, but also differences, with that performed in \cite{Brida2011}. In both cases, experimental variance estimators are shown to include deterministic terms. However, these terms concern in \cite{Brida2011} variation of quantum efficiency from a pixel to another, in a regime of proportional detection of fluences of several photons per pixel by a conventional CCD. In the photon-counting regime, the thresholding procedure rends this term negligible. Hence, the corrections affect both the variance  and the variance of the difference in \cite{Brida2011}, while our correction consists in subtracting from the variance and from the corrected mean a deterministic term coming from the global gaussian shape of the SPDC, with no modification of the variance of the difference, in order to avoid the measurement of a quantum regime ($F(S_I)<1)$ in Eq.\ref{fdeltarvmoins}) for independent images.

 \section{Experimental results}\label{experiment}

\begin{figure}
\centering\includegraphics[width=1.00\columnwidth]{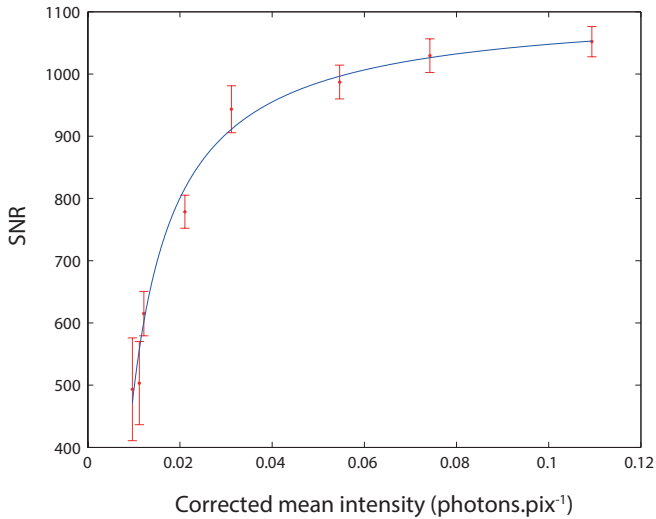}
\caption{\label{fig:1} Color online. SNR versus the corrected mean intensity for 700 images and 384*384 pixels. Red stars: experimental values. Blue curve: Eq. \ref{SNRexp} with parameters given in the text. The errorbars are 95 $\%$ confidence intervals deduced from the standard deviation of the SNR for sets of 10 series of 70 images.}\label{SNRfig}
\end{figure}

 Fig \ref{SNRfig} shows the experimental signal-to-noise ratio of the mean intercorrelation of couples of far-field signal-idler images of SPDC, for different intensity levels. The experiment involves two cameras and details of the experimental set-up can be found in \cite{Moreau2014b}. To calculate this SNR, we first fit the intercorrelation peak with a gaussian function, then sum the values of this function for all pixels. The term of noise is directly given by the standard deviation of the mean intercorrelation in an area where its mean value is zero, i.e far from the intercorrelation peak. With this experimental procedure, Eq \ref{SNR} of our idealized model remains valid even for a coherence cell not reduced to one pixel (note however that this method is applicable only if the position of the correlation peak has been previously determined). Indeed, the theoretical value issued  from this equation appears to be in good agreement with the experimental values. The measured quantity is the sum of the photoelectrons and of the noise $I=m+p_{n}$. By applying the correction proposed in subsection \ref{satur}, Eq \ref{SNR} becomes:

 \begin{equation}\label{SNRexp}
SNR\simeq \sqrt{D\ K}\frac{\eta\  (I(1-I))-p_{n}}{I(1-I)}
\end{equation}

With D=384x384 pixels and K=700 images, the best fit of the experimental points of Fig \ref{SNRfig}, using a standard nonlinear least-squares procedure, is obtained with $\eta=0.109$ and $p_{n}=5.6\ 10^{-3}$. This latter value is compatible with the usual level of CIC in an EMCCD camera \cite{Lantz2008}. Moreover, the standard deviation of the mean intercorrelation function appears, for all intensity levels, to be equal within 10\% to $I(1-I)/ \sqrt{D\ K}$. On the other hand, $\eta=0.109$ does not correspond to the actual overall quantum efficiency of the system, which is approximately 0.5 by taking into account all the optical elements. A part of the discrepancy could come from stray light generated by fluorescence of the optical elements. Because the level of this fluorescence is proportional to the pump intensity, like SPDC and unlike CIC, its effect is similar to a decrease of the quantum efficiency. Indeed, if $F$ photons coming from a pair plus $\alpha F$ single photons are incident on one of the twin pixels of our idealized model, the mean number of photoelectrons becomes $m=\eta(1+\alpha)F$, while the correlation signal remains equal to $\eta^{2} F=\eta' m$, with $\eta'=\eta/(1+\alpha)$. Nevertheless, even by taking into account this fluorescence $\eta=0.109$ is weaker than expected. Note that the degree of correlation can be directly assessed from Eq.\ref{fs2} or \ref{fdeltarvmoins}, with the same too weak values ~\cite{Devaux2012}. As detailed in \cite{Meda2014}, many geometrical factors can affect the coefficient $F(S_I)$ of Eq. \ref{fdeltarvmoins}. In short, if the signal and idler surfaces are not in strict correspondence, the variance of the difference increases. However, we do not believe that the low value of $\eta$ given by our fitting procedure is related to such geometrical aspects. Indeed, if both photons of a pair are detected, they increase the integral of the correlation peak even if they are not detected on the right pixels. Hence, geometrical imperfections would result at best in a shifting of the correlation peak, at worst to an enlargement of the peak, but not to loss of photons. Since there is a good correspondence between the integral of the normalized intercorrelation function (Eq.\ref{fs2}) and the variance of the difference for sufficiently binned pixels (Eq. \ref{fdeltarvmoins}), we think that the geometrical correspondence is correct, resulting also in a correlation peak whose lateral dimensions are reasonably close of the theoretical values \cite{Moreau2014b}.
								
\section{Conclusion}

We have shown that the amount of quantum correlation between twin images can be determined by assessing the signal-to noise ratio of the intercorrelation signal. This SNR scales as the square root of the number of involved coherence cells and images. Hence, it seems possible to obtain results on single couples of images for a sufficient number of coherence cells in the image. The SNR increases also with the mean level in the image, inducing the necessity of working at relatively high level, unlike with the visibility criterion. Furthermore, the quantum nature of the correlation must be proved by computing  the variance of the difference, that should be subpoissonnian after correction of two artifacts, that both could lead to an apparent subpoissonnian correlation of images that are not correlated at all. The first artifact is due to the non negligible probability of multiple photons on one pixel. The second is related to the deterministic spatial profile of the SPDC.

\section{Acknowledgments}
This work has been partly supported by the Labex ACTION program (contract ANR-11-LABX-0001-01).

\section{Appendix}
\begin{figure}
\centering\includegraphics[width=1.00\columnwidth]{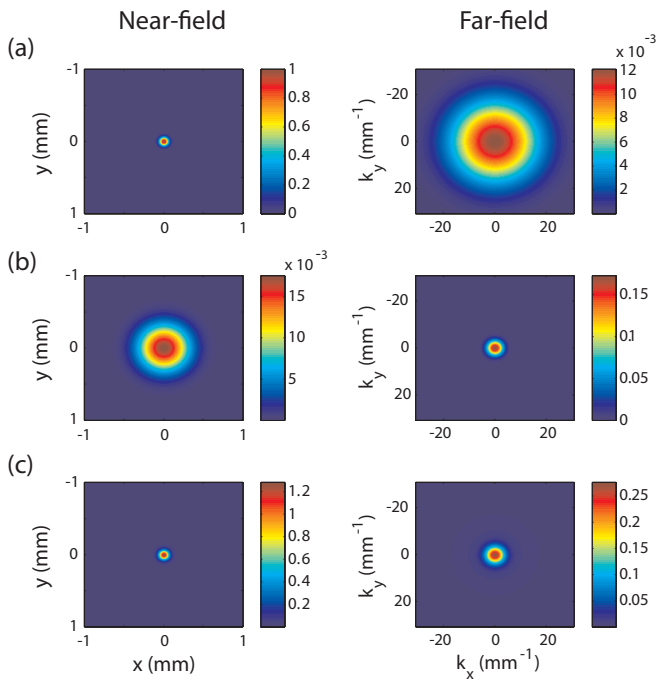}
\caption{\label{fig:2} Color online. Intensity profiles of the beams in the near and far field.  (a) Gaussian beam with a waist of $w_0=0.0847 mm$.(b) Gaussian beam with a waist of $w_0=0.4525\ mm$. (c) Beam resulting of the coherent sum of the two previous beams. Colorbars correspond to intensity levels expressed in arbitrary units.}
\end{figure}

Here we show how a classical beam of light can appears to violate the Heisenberg uncertainty principle if measured with a one dimensional detector. Such a classical state can be built by combining coherently two Gaussian beams. As shown on fig. \ref{fig:2}, we add two Gaussian beams of different sizes (a) and (b) to build the state presented in (c).
\begin{figure}
\centering\includegraphics[width=1.00\columnwidth]{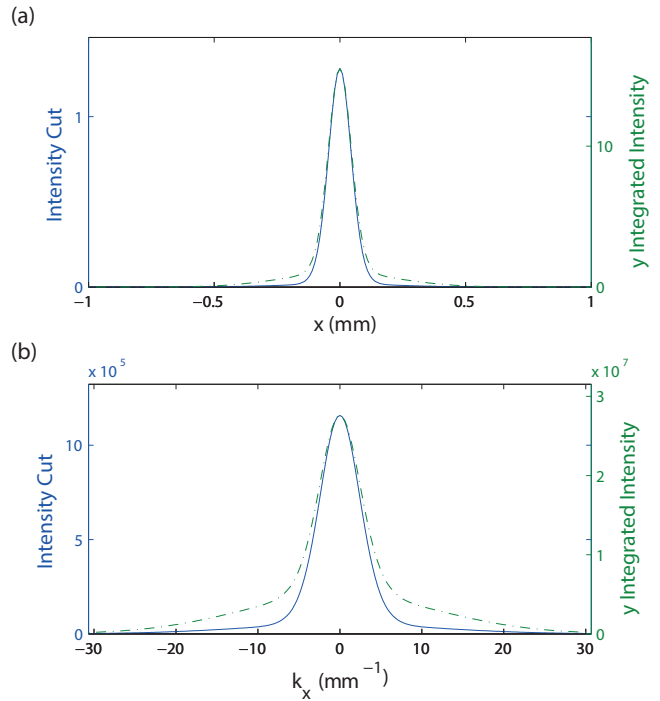}
\caption{\label{fig:3} Color online. Full blue lines: one dimensional intensity profiles which correspond to the selection of the row of pixels $y=0$ in both the near-field (a) and far-field (b). Intensity levels are expressed in arbitrary units. Dashed green lines: corresponding y integrated intensities. }
\end{figure}
If now we apply a one dimensional detection on this last state, by selecting a row of pixels in the intensity beam profiles in both the near and the far field, we obtain the distributions presented on fig. \ref{fig:3}. We can then evaluate the standard deviation of those intensity distribution curves, in order to test the Heisenberg uncertainty principle. In our example, assuming a pixel size of $10\ \mu m$ in the near field, we obtain the standard deviation in the near field $\sigma_x=0.0733\ mm$ and in the far-field $\sigma_p=5.51\hbar\ mm^{-1}$, giving a product of
\begin{equation}
\sigma_x\sigma_p=0.4039\hbar<\frac{\hbar}{2}
\end{equation}
This equation exhibits an apparent violation of the Heisenberg inequality. Of course the Heisenberg uncertainty principle is in fact verified by the state that we have built classically in fig. \ref{fig:2}(c). This false violation appears because of the lack of consistency in the previous test of the Heisenberg inequalities. Indeed, by applying a one dimensional detection we are led to evaluate $\sigma_x$ and $\sigma_p$ on two distinct subsystems. As a consequence, the product of the standard deviations has no longer meaning because the Heisenberg principle governs the behavior of an unique system. We can easily be convinced that the two involved subsystem in the previous test were not the same, since the subsystem that is selected by 1-D detection in the near field would diffract in the far field on the whole plane after passing through a one pixel wide slit. The system involved in the far field, where detection is also performed on a row of pixels, is clearly different. Actually, at the center of the beam the contribution of the smallest Gaussian will always be favored in both the near and far field, leading to the apparent violation, while the largest beam is diluted in two dimensions because of the intrinsic two dimensionality of the diffraction phenomenon. For the same reasons, away from the center the most important contribution is always due to the larger Gaussians and those contributions have to be taken into account to maintain the consistency of the demonstration.\\
In order to maintain the consistency, the evaluations of $\sigma_x$ and $\sigma_p$ have to be done on the whole system by integrating on the two spatial transverse dimensions. By integrating on y for the light system presented in fig. \ref{fig:2}(c), we obtain the green dashed lines on fig. \ref{fig:3}, with a standard deviation $\sigma_x=0.12\ mm$ in the near-field and $\sigma_p=8.54\hbar\ mm^{-1}$ in the far-field, giving the product
\begin{equation}
\sigma_x\sigma_p=1.029\hbar
\end{equation}
thus satisfying the Heisenberg inequality as expected for classical light.\\
It should be noted that evidencing an EPR paradox requires the demonstration of the sub-Heisenberg behavior of correlations. As a consequence, the demonstration has to be done in the same context as that in which the Heisenberg principle can be correctly tested. In particular, such an EPR demonstration has to satisfy the criterion of uniqueness of the quantum system. The whole light system has therefore to be involved in both the near-field and far-field and only a two dimensional integration would be consistent.

%\bibliography{SNRbiblio}

\end{document}